\documentclass[12pt]{article}
\usepackage{color,multicol}
\usepackage{amsmath}
\usepackage{amssymb}
\usepackage{epsfig}
\usepackage{pifont}
\usepackage{graphicx}
\usepackage{epstopdf}

\usepackage{tocloft}

\begin{document}

\title{Why Quantitative Structuring?\footnote{Original version 25 July 2015, this presentation 7 September 2020.\vspace*{1mm}}}

\author{Andrei N. Soklakov\footnote{Head of Strategic Development, Asia-Pacific Equities, Deutsche Bank.\newline
{\sl The views expressed herein should not be considered as investment advice or promotion. They represent personal research of the author and do not necessarily reflect the view of his employers, or their associates or affiliates.} Andrei.Soklakov@(db.com, gmail.com).}}
\date{}
\maketitle

%\vspace*{-0.5cm}

\begin{center}
%25 July 2015\\[1cm]
\parbox{12.5cm}{
{\small
Quality-designed consumer products are easy to recognize. Wouldn't it be great if the quality of financial products became just as apparent?\\

This paper is addressed to financial practitioners. It provides an informal introduction to Quantitative Structuring -- a technology of manufacturing quality financial products (information derivatives).\\

The presentation is arranged in three parts: the main text assumes no prior knowledge of the topic; important detailed discussions are arranged as a set of appendices; finally, a list of references provides further details including applications beyond product design: from model risk to economics and statistics.
}
}
\end{center}

\vspace*{5mm}

\newpage

\section{Motivation}

We believe that products are very important for any business. Products pay our salaries and define our relationships with clients. A business without a product is a business in trouble. Successful products open many doors -- anything from business expansion to the support of charitable causes becomes possible. Products are quite simply the ultimate source of our risks and rewards.

We believe that every effort must be made to improve financial products. Almost every other industry has seen dramatic improvements in product design. The financial industry cannot afford staying behind.

We believe that the future of finance lies in adopting a more scientific approach to product design. We note that the numerous successful applications of science in other industries became possible only because science itself reached the necessary level of clarity. To achieve similar results in finance, we should be prepared to upgrade the scientific standing of finance.

We expect many challenges along the way. It will help us to remember that there were tough times in the history of every single branch of science. During such times problems appear to be special, even beyond the realm of logic. Moral dilemmas, regulatory pressures and the need to persuade the public dominate the agenda. Somehow, these are also the times when progress is made. Right now is probably the best time ever to discuss improvements in financial products.

\section{Guiding principle}\label{sec:GuidingPrinciples}
Quality research is hard. To achieve it we have to fight our own prejudice and numerous subconscious biases. This is what science is really all about. Science is not confined to any particular subject, it happens whenever there is a genuine and honest attempt to understand something.

We want to understand how to make \lq\lq good'' financial products. Clearly, not every product is \lq\lq good''. There must be {\nobreak properties}, fundamental laws if you will, which \lq\lq good'' financial products must satisfy. But how do we look for these fundamental laws? How do we even know if we have found one? This is an extremely hard question. We need some inspiration, and there is no better place for finding that than in the real success stories of science.

Let us pick some well-understood scientific theories which are as different from each other as possible. Newtonian Mechanics and Darwinian Evolution are good examples. What do these theories have in common? What makes them useful? Why do we teach these theories to our children despite well known factual contradictions?

Both theories contain a powerful observation, a paradigm which greatly simplifies and facilitates understanding. For example, the famous Newton's law, $F=ma$, is just a definition of a quantity which Newton decided to call \lq\lq force". Mathematically speaking, it is a triviality. So, where is the breakthrough, where is the insight? -- we might ask. The insight lies in the fact that thinking in terms of forces greatly simplifies our understanding of many physical phenomena. So much so, that the greatest achievements of Newton are now accessible to school children. Similarly, Darwin's observation of natural selection gives us a thinking paradigm, a concept. This concept is not even quantitative, but it makes the living world much easier to understand.

So, this is what we need to find -- {\it a technical concept which makes financial products easier to understand}.

[ If we had to jump ahead and reveal our candidate for this magical concept we would say that good financial products are always based on research, and the optimal products are most easily understood via {\it likelihood functions} describing this research (Sec. 4). ]

\section{Financial products}\label{sec:FinancialProducts}
What exactly are financial products? Browsing through termsheets we quickly discover that each and every product is really just a function, $F(x)$, which states how benefits (normally cashflows) depend on the underlying variables, $x$ (which may include time, market prices, credit ratings, weather readings or actions of other people -- anything that is relevant for a given product). In the following we refer to $F(x)$ as a payoff function.

Now we know what our theory has to produce -- payoff functions.

The landscape of all financial products is huge. We need a good place to start the exploration. In Appendix~\ref{sec:WhyInvestments} we consider all possibilities and conclude that investment products provide a very good starting point.

\subsection{Investment products}\label{sec:InvestmentProducts}
Investment structuring is an old problem. Even Modern Portfolio Theory is now over 60 years old. Nevertheless, the quality of investment products still has room for improvement. Analysis which leads to this conclusion is presented in Appendix~\ref{sec:LegacyIssues}. Here we focus on the constructive elements and ask ourselves what minimal features we want to see in a good investment product.

We demand that each investment product has a well-defined purpose, accurately expresses clients' views and has logical integrity. These three requirements are not independent. Let us examine them as we would examine the facets of a crystal when looking for the most promising direction of study.

\vspace*{3mm}
\ding{172} {\bf Purpose}\\
Each investment product must have a goal. Mathematically, this is formulated as an optimization problem. Although routinely violated in practice (see Appendix~\ref{sec:LegacyIssues}), this requirement is very well established in finance and economics. The best known example is probably the Markowitz optimal portfolio which is constructed as a solution to a particular mean-variance optimization problem. Expected utility theory provides a more general framework for rational investors. Even clients who choose to depart from rationality have some means of describing their goals -- the optimization setup of prospect theory\footnote{The relationship between our approach and Behavioral Finance is outlined in Appendix~\ref{sec:BehavioralFinance}.}.\\

\ding{173} {\bf Accurately expressed views}\\
Investors may agree with the market, but most often they do not. Investors search for undervalued or overvalued opportunities and bring this new information to the market. It is very important for both the investor and wider society that the results of their research are expressed accurately. This means that investment products must be able to reflect subtle differences in investors' views. Investors should be able to combine views and there should be no implicit extrapolation of views beyond the researched scope. Further clarifications of these requirements can be found in Appendix~\ref{sec:LegacyIssues}.

\ding{174} {\bf Logical integrity}\\
Information processing is a big part of investment activity. Information processing obeys laws commonly known as logic. Even before we explore what the logical integrity of an investment looks like we know that such integrity must be important.

\vspace*{3mm}
The first of the above requirements has been extensively explored within the field of economics. The second requirement is obvious in that it should follow from any reasonable approach to investment. The third requirement is relatively new and it requires major clarification. In the next section we explain what we mean by logical integrity via its implementation. Because logic is the backbone of any branch of science, we hope that this relatively new line of enquiry will give us a glimpse of our ultimate dream -- finance as a scientific discipline.

\section{Logic, likelihoods and information derivatives}
Let $x$ be some underlying variable (as introduced above in section~\ref{sec:FinancialProducts}). In the presence of uncertainty, our knowledge about $x$ is described by a probability distribution $p(x)$. Upon learning new data, $d$, knowledge $p(x)$ should be updated to $p(x|d)$. The logic behind this update is well known in probability theory as the Bayes' theorem~\cite{Jaynes_2003}. This reads
\begin{equation}\label{Eq:Bayes}
p(x|d)={\cal L}_d(x)\,p(x)\,,
\end{equation}
where ${\cal L}_d(x)$ is called the {\it likelihood function}. As we can see from the above equation the likelihood function encapsulates everything we need to know in order to update our knowledge about $x$ on the account of learning $d$.

Imagine an investor for whom we can reconstruct the entire logical path. This means that we know the starting distribution $p(x)$ assumed by the investor and all the subsequent learning steps of the form~(\ref{Eq:Bayes}) which lead to the final knowledge
\begin{equation}\label{Eq:LogicalPath}
p(x|d_1,d_2,...,d_n)={\cal L}_{d_n}(x)\cdot\dots\cdot{\cal L}_{d_2}(x){\cal L}_{d_1}(x)\,p(x)\,.
\end{equation}
It is important that the starting distribution, $p(x)$, also known as the prior, is built by the investor using publicly available (market) information and that the data $d_1,d_2,...,d_n$ (which may include assumptions as well as established facts) come in addition to this prior knowledge. In \cite{Soklakov_2013b} we proposed to call such investors {\it logical}.

\newpage

The concept of a likelihood function is key in describing logical investors. Earlier we noted that financial products are also defined by functions -- the payoff functions. We see that within any sensible investment theory the two concepts -- the likelihood and the payoff functions -- must be connected.

Historically, we have not had much science about payoff functions. Typically, the structure of financial derivatives has been either postulated or assumed as given -- suggested spontaneously by wise markets. By contrast, the concept of a likelihood function is extremely well known across all kinds of scientific applications: probability and statistics, computer science and artificial intelligence, physics and engineering, biology and medicine -- to name just a few.

It should come as no surprise that thinking in terms of likelihood functions greatly simplifies the understanding and structuring of investment products. Mathematically, this manifests in the form of very simple intuitive equations for payoff functions (see Appendix~\ref{sec:Equations}).

Thinking in terms of likelihoods makes our products explicitly dependent on all the information included in the relevant learning step. For this reason we propose to call the resulting products {\it information derivatives}.\footnote{This generalizes our earlier definition proposed in~\cite{Soklakov_2008}.}

\section{Results so far}

\subsection{Theory}
The main technical result is a pair of very simple equations (see Appendix~\ref{sec:Equations}). These equations allow us to design products which satisfy all of the requirements discussed in this paper. In particular, going through the key requirements from section~\ref{sec:InvestmentProducts}, we find:
\begin{itemize}
 \item[\ding{172}:] Each of our products has a well-defined goal. If required, this goal can be written as an optimization problem in the notation of the expected utility theory.

     For the first time, we can design a financial derivative and claim it to be \lq\lq the best product" not just because it is a good marketing move but
     because it is true (in a very well-defined mathematical and economic sense).

 \item[\ding{173}:] All of our products accurately express clients' views. The views of the client (or their research advisor) are included in a transparent way as inputs of our equations.

     The equations are very simple so they can also be used in reverse: to show the views expressed by the clients' current positions; or the views that the client would be expressing if they went for any particular product~\cite{Soklakov_2011}. This supports a meaningful and constructive conversation with the client making sure that they do end up expressing the views that they actually have.

 \item[\ding{174}:] The logical integrity of new products is ensured by the connection of the payoff functions with the likelihood functions. This connection is woven in the derivation of the equations~\cite{Soklakov_2013b, Soklakov_2011} and is the reason for their simplicity and intuitive convenience.
\end{itemize}

\subsection{Practical and scientific considerations}

With all their shortcomings, financial products of the past have given us a tremendous amount of experience. We respect this knowledge and want to retain as much of it as possible.\footnote{In science this approach is known as the {\it correspondence principle}. It ensures accumulation of scientific knowledge. Every new theory is required to retain all of the useful (valid) insights from its predecessors.} To this end we require our theory to have the ability to analyze traditional products even though it may expose them as deficient in some way. In~\cite{Soklakov_2013a} we check this requirement by looking at some traditional products. We start with vanilla instruments (spot, vol, skew products) moving into more exotic path-dependent derivatives. We also reproduce key design ideas of bespoke equity indices and discuss the early exercise feature. As expected, we find that traditional products use hidden assumptions which often make some approximate sense but in general may not be easy to justify.

In the same paper~\cite{Soklakov_2013a} we noticed that product and model design form related disciplines. Indeed, modeling deficiencies can be exploited as investment opportunities. We develop this idea further in~\cite{Soklakov_2014MR} where we present an economically meaningful approach to model risk assessment.

In Ref~\cite{Soklakov_2013b} we focused our attention on risk aversion. This very important topic arises in practice every time we try to be conservative or stay on the safer side of some investment strategy. Using our equations we found that the standard ad-hoc methods of engineering risk aversion (such as expressing a more conservative view or modifying payoffs) do not normally achieve their goal. Instead, investments can very easily turn into a gamble. Being conservative turns out to be a delicate task and our equations provide the tools for accomplishing this task.

In Ref~\cite{Soklakov_2014EqPuzzle} we touch on the subject of long-term investments. We challenge our theory by considering the equity premium puzzle which has been defying mainstream economics for the last 30 years. In the long run, riskier assets (such as equities) tend to produce higher returns than the less risky ones (such as bonds). What is puzzling is the numerical magnitude of the effect: the standard (consumption-based) economic theory predicts a much smaller effect (an order of magnitude smaller) than observed in the market. Resolving the puzzle is considered important because of its relevance to socially important long-term investments such as pensions. With just a few lines of mathematics our theory predicts a correct ballpark figure for the equity premium. In the context of the scientific method this result is especially important because it tests the validity of our approach against real data.

Ironically, innovation often struggles to explain the benefits it brings. This can be a major practical challenge, even when the target audience are experts. Take, for example, the light bulb. Today it is used as an icon symbolizing all ideas and innovation. In 1878 a British Parliament Committee described Edison's light bulb as {\it \lq\lq good enough for our Transatlantic friends ... but unworthy of the attention of practical or scientific men''}. In 1911, speaking in his role as Professor of Strategy (at Ecole Superieure de Guerre) the famous French general, military strategist and later World War I Allied Commander-in-Chief Ferdinand Foch described airplanes as {\it \lq\lq interesting toys ... of no military value''}. In 1943, Thomas Watson, chairman of IBM, talking about the commercial future of computing estimated: {\it \lq\lq I think there is a world market for maybe five computers''}.\\

Prejudice against innovation is clearly a strong and important phenomenon which poses a major practical problem. Within the scientific approach there is only one thing we can do about that. Confronted by misunderstanding or prejudice we should research its origins. In doing so we might find a way of addressing cognitive difficulties, improve our products or even discover a very different set of needs.

Thinking in terms of likelihood functions, we see that financial products are intimately connected to probabilities. Perhaps some of our difficulties mirror the general lack of intuition regarding probabilistic and statistical concepts.

Indeed, we show~\cite{Soklakov_2018} that understanding financial performance of information derivatives is very helpful for understanding disagreement between probability distributions. In statistics such disagreements are quantified by highly abstract (axiomatically motivated) measures called divergencies. Our financial intuition thus helps statistical understanding.

We further show how understanding statistical disagreements, in turn, creates trading opportunities~\cite{Soklakov_2018}. Indeed, imagine a couple of people who disagree on probabilities for some future event. This is enough for them to trade. Indeed, because they believe different probability distributions they would arrive at different expectations. As a result, one can design a zero-sum trade which, in terms of expectations, is likely to look attractive to both people. In effect, we now know how to structure not just individual products but entire markets (where investors with different views make a market for each other, see Ref.~\cite{Soklakov_2018} and Appendix V of Ref.~\cite{Soklakov_2014EqPuzzle}).

Having committed to the scientific path in understanding financial products we cannot escape the fact that financial planning is a physical process which happens inside the human brain. In Section 4.3 of~\cite{Soklakov_2018} we reconciled our structuring equations with the results of independent neurophysiological experiments. This, of course, is only the beginning. As experimental studies of the brain intensify, we expect a lot more data to become available. Our theory will evolve accordingly. We imagine a future in which really successful financial products are built as reflections or extensions of popular cognitive strategies implemented by the brain.

\section{Appendix: FAQs and discussions}

\subsection{Why so much focus on investments?}\label{sec:WhyInvestments}

Let us ask ourselves a more fundamental question: Why do we have customers? People come to us with different stories but all of them point to three fundamental reasons: investment, raising capital and hedging. On top of these we must remember requirements from wider society, which can be summarized as providing efficient allocation of resources wherever needed.

Thinking through the above reasons and requirements, we see that investment products are taking a lead role. Indeed, investment and raising capital are really two sides of the same coin: a customer who is trying to raise capital is looking for investors and must therefore present them with an investable opportunity.

Even when we discuss solutions to social issues we talk about investments: in schools, in hospitals, in local communities. The benefits of such investments may be difficult to capture mathematically, but there is no doubt -- efficient allocation of resources is an investment-type problem.

Worried about older people? Well, pensions are examples of long-term investments. In fact, any kind of serious economic planning needs good quality investment behavior, for it is the prime supplier of accurate market information.

Right now our trust in investment products is at an all time low. Regulation is pouring over investment banks like concrete over leaking nuclear reactors. At the same time we understand that outlawing investment products is not an option: in addition to all of the above reasons, this will encourage speculation in products which have never been designed as investment vehicles: debt insurance and other hedging instruments. Improving the design of investment products appears to be an urgent priority.

\subsection{Legacy investment products -- what are the issues?}\label{sec:LegacyIssues}

Most people would agree that our investment banks needs serious improvement. However, it is always a good exercise to identify issues as precisely as possible. In this section we summarize our argument for reform of investment products.

We start by looking at the most basic promise of each and every investment product -- the promise to express accurately investors' views relative to the market. Let us examine the quality of this promise. The following points summarize three types of problems which we regard as critical.

{\bf View differentiation}\\
Consider two investors with views similar in direction but very different in strength. Imagine a market of options with a well pronounced skew. One investor believes that the skew should be a bit less pronounced than the market-implied while the other thinks that there should be no skew at all. The difference in views is significant, but because both investors agree on the direction, currently they would be offered the same set of products.

{\bf View integration}\\
Consider an investor with a view on both the volatility and the skew. The investor believes that the skew should disappear and the volatility will realize 5 vol points below the current market expectation. How should the investor allocate their money? Should they put most of it on the volatility or on the skew? Are we even sure that the investor is better off with two separate trades (one on the vol and the other on the skew), or is there a structured product which is better optimized to express the combined view? Structurers are currently poorly equipped to tackle these types of questions.

The above two problems are already pretty bad. We cannot differentiate or combine customer views! An equivalent situation in mathematics or science would be like having a number theory which does not tell us how to add numbers or a physical theory which cannot capture the relative strength of forces. Surely, the situation cannot get any worse than this. The following observation says that it actually does. \\ \\

{\bf View extrapolation}\\
Investment products often extrapolate customer views beyond their original research context. Even if the expressed view makes sense in a limited region (e.g. near ATM), the extrapolated view often has no bounds. Indeed, quite a few investment products (including standardized derivatives) offer theoretically unlimited gains or can lead to theoretically unlimited losses -- all mostly in the circumstances which are very difficult to research (no market information, low-probability scenarios, etc.). Such extrapolation of views can lead to systemic accumulation of completely unnecessary risks which are often in the tails.

It would be useful if in addition to the above, we could also give specific examples pointing to the flaws of real-life products. The presentations by Merton~\cite{Merton_MIT} and Dupire~\cite{Dupire_2011} are good independent examples of such analysis for simple and complex derivatives respectively.

However, as soon as we start thinking about such examples we quickly realize that no finite number of them is ever sufficient -- just like debunking one perpetual motion machine after another does not stop all attempts to invent new ones. How could we show that investment products contain more than a few rotten apples, that huge classes of them are fundamentally flawed? There is only one way of doing that -- we need a method, an ability to create any number of realistic examples on demand. Here is one such method...

The first thing we need to acknowledge is that any investment is an optimization. This is in fact true in a much wider range of situations -- pretty much anything that we \lq\lq want'' anything that has a \lq\lq goal'' can be viewed as an optimization. The language of maximizing returns, minimizing risks and constraining losses is the language of optimization which investors use naturally to describe their goals.

To make it as fair as possible, let us use some examples from readers' personal experience. So please sit back, relax, and remember as many investment products as you can. Barrier options, lookback structures, mountain ranges, cliquets -- the whole lot. Take as much time as you like. And as you are recalling them, choose you favorite product. Make sure you are really comfortable with your choice. Imagine all of the features of this product. Remember all the clever ideas behind these individual features. Now, could you please explain: what optimization problem does it solve? I don't know which product you have chosen but I am pretty sure that you are now struggling with this question. I hear some people trying to talk about bears and bulls, but let us remain focused: What optimization problem does it solve? Animal spirits are not helping. The more you think the more you realize that, strictly speaking, the product was never designed to solve any optimization. In fact, nobody even bothered to state the problem mathematically.

This could not possibly be good -- neither for the client nor for wider society. There are of course exceptions to the above argument, but we all know that the vast majority of investment products would not stand up to this very simple examination.

The good news is that all of the specific issues discussed in this section can be resolved by Quantitative Structuring (see the main text).

\subsection{Structuring equations}\label{sec:Equations}
In Ref.~\cite{Soklakov_2013b} we discovered that for a large class of logical investors, the logical path~(\ref{Eq:LogicalPath}) can be viewed as a simple two-step learning process: one step incorporating market research and the other learning investors' private information (risk aversion). The two learning steps gave us two equations which, by way of introduction, we summarize here as
\begin{equation}\label{Eq:b=fm}
b=f\,m
\end{equation}
\begin{equation}\label{Eq:PayoffElasticity}
\frac{d\,\ln F}{d\,\ln f}=\frac{1}{R}\,.
\end{equation}
One way of explaining these equations is to note that they are obeyed by a payoff function $F(x)$ which solves the following optimization problem
\begin{equation}\label{Eq:Optimization}
\max_F\int b(x) U(F(x))\,dx\ \ \ \ {\rm subject\ to\ budget\ constraint}\ \ \ \ \int F(x) m(x)\,dx=1\,.
\end{equation}
Risk aversion coefficient $R$ is connected to the utility $U$ through the standard Arrow-Pratt formula: $R=-FU''_{FF}/U'_F$. The economic meaning of the market-implied and investor-believed distributions $m(x)$ and $b(x)$ is explained by the above optimization problem.

The connection between equations~(\ref{Eq:b=fm}) and (\ref{Eq:PayoffElasticity}) to the optimization~(\ref{Eq:Optimization}) shows that all our investment products have a well-defined economic goal which is consistent with the expected utility theory. The first requirement of section~\ref{sec:InvestmentProducts} is satisfied.

There is of course more to our equations than their connection to optimization problem~(\ref{Eq:Optimization}). Coming back to the structure of logical paths~(\ref{Eq:LogicalPath}), we note that equivalent conclusions can be reached via many different logical paths. Consequently, there are many equivalent forms our equations can take. These forms differ in the intermediate payoff functions, such as the function $f(x)$ which is typically found from Eq.~(\ref{Eq:b=fm}) and then used to obtain the final payoff function $F$ by integrating Eq.~(\ref{Eq:PayoffElasticity}).

The intermediate function $f(x)$ in our equations has the meaning of the optimal payoff function of the growth-optimizing investor -- the investor which aims for the greatest expected rate of return. For such investor $R=1$, Eq.~(\ref{Eq:PayoffElasticity}) becomes redundant and we are only left with Eq.~(\ref{Eq:b=fm}).

Looking closely at Eq.~(\ref{Eq:b=fm}) we see that it has an easily-recognizable Bayesian structure~(\ref{Eq:Bayes}) with the payoff $f$ playing the role of the likelihood function~\cite{Soklakov_2011}. For the growth-optimizing investor the connection between payoff and likelihood functions takes its simplest form -- they simply coincide. Thinking in terms of likelihood functions is in many ways equivalent to thinking in terms of growth-optimizing investors.

First introduced by Bernoulli over 270 years ago, the concept of a growth-optimizing investor is widely researched and used. Economists use it every time they mention log utility, the hedge fund community know it under the name of Kelly strategy. Promoted by some and criticized by others, growth-optimizing strategy left very few researchers without a strong opinion on it. This includes legendary fund managers as well as Nobel prize economists. Paul Samuelson left us with a unique illustration of how far people are prepared to go to specify their own position relative to the growth-optimizing strategy. Samuelson decided to address his audience with a paper which was painstakingly edited to consist of exclusively monosyllabic words~\cite{Samuelson_1979}. Our morally fragile industry finally had a leader which was prepared to insult his own audience on an almost poetic level (not to mention the epic sacrifice of clarity in such a presentation).

With the huge benefit of hindsight and with deepest respect to our pioneers (including their mistakes), we note that the heated debates of the past missed a very important point. The readiness with which people position themselves relative to the growth-optimizing investor and the strength of their opinions makes the growth-optimizing investor a very good benchmark of investment behavior -- a very convenient intermediate step. Likelihood-based thinking exposes this fact, benefits from it and provides striking conceptual and technical simplicity (just think of understanding and solving Eqs.~(\ref{Eq:b=fm}) and~(\ref{Eq:PayoffElasticity})).

\subsection{Behavioral Finance}\label{sec:BehavioralFinance}
One of the frequently asked questions on Quantitative Structuring is how our approach (which is rational by construction) can coexist with the findings of Behavioral Finance. This is a very interesting question. In fact this is much more than just a question -- this is an opening to a very important debate about how we deal with the imperfections of human judgement. On the one hand we must be aware and respectful of such limitations (they are real measurable psychological phenomena), and on the other hand we must help people to overcome their limitations.

The best way of clarifying our current position in this debate is to give an example. Imagine a questionnaire on an established technical subject. To be specific, let us consider a test in geometry. Imagine that a wide representative set of people participated in this test. What results are we going to see and what conclusions should we make from that?

Pretty clearly, we are going to see people making mistakes. There will be patterns in our data -- some mistakes will be more frequent than others. Some of the mistakes may be very robust and even traceable to various behavioral and cognitive tendencies (just think of optical illusions). The knowledge of this can be very important: humans constantly interpret geometrical configurations (such as judging distances) in the context of their behavior (such as driving a car).

So what shall we do about this rich database of mistakes? Shall we elevate them to the rank of fundamental laws of nature and replace mathematical fact with \lq\lq behavioral geometry"? No. Of course not. What we should do is build tools that help people to overcome their limitations (e.g. chevrons on the motorways that help to judge distances, pocket calculators to assist with mental arithmetic, etc.).

In the above example, we refused to override the axioms of geometry on the account of human error, but at the same time we acknowledged the importance of cognitive effects. There is no contradiction there. Similarly, we should not contrast Behavioral Finance with the need to facilitate rational behavior among people. The two go hand in hand. Both approaches are based on healthy scepticism about the efficiency of the markets. In Quantitative Structuring we no longer trust the markets to come up (spontaneously) with sensible financial derivatives -- we decide to build tools to help us with our cognitive limitations.

\end{document}